\pdfoutput=1
\documentclass[10pt,a4paper,fleqn,british,hyphens]{elsarticle}
\usepackage[T1]{fontenc}
\usepackage[utf8]{inputenc}
\usepackage{color}
\usepackage{babel}
\usepackage{float}
\usepackage{amsmath}
\usepackage{graphicx}
\usepackage{rotating}
\usepackage{rotfloat}
\usepackage{microtype}
\usepackage[unicode=true,
 bookmarks=false,
 breaklinks=false,pdfborder={0 0 1},backref=false,colorlinks=true]
 {hyperref}
\hypersetup{
 urlcolor=blue, citecolor=magenta}

\makeatletter

\pdfpageheight\paperheight
\pdfpagewidth\paperwidth

\providecommand{\tabularnewline}{\\}

\@ifundefined{date}{}{\date{}}

\usepackage{colortbl}

\usepackage{pdfsync}

\makeatother

\begin{document}

\begin{frontmatter}{}

\title{{\large{}CovidChain: An Anonymity Preserving Blockchain Based Framework for Protection Against Covid-19}}

\author[cua]{Hiten Choudhury}

\ead{hiten.choudhury@cottonuniversity.ac.in}

\author[cua]{Bidisha Goswami }

\ead{bidishagoswami112@gmail.com}

\author[smc]{Sameer Kumar Gurung\corref{cor1}}

\ead{skg@smcs.ac.in}

\cortext[cor1]{Corresponding Author }

\address[cua]{Department of Computer Science \& IT,Cotton University, Guwahati, Assam,
India.}

\address[smc]{Department of Computer Science, Saint Mary's College, Shillong, Meghalaya,
India.}
\begin{abstract}
Today, the entire world is facing incredible health and economic challenges
due to the rapid spread of the life threatening novel Coronavirus
Disease - 2019 (COVID-19). In the prevailing situation when a vaccine
is many months away, the way forward seems to be a controlled exit
from the lockdown - where, infected/exposed people are strictly quarantined
and recovered/unexposed people are allowed to carry on with their
day to day business activities. However, appropriate physical distancing
norms will have to be strictly followed for such relaxations. Therefore,
mechanisms are required that will assist people in following the social
and physical distancing norms in public places. In this paper, we
propose an anonymity preserving blockchain based framework that allows people, through use
of their smart phones and other communication devices, to protect
themselves from infections as they conduct their daily business activities.
\end{abstract}
\begin{keyword}
Covid-19, Coronavirus, Contact Tracing, Blockchain, Security, Privacy, Anonymity.
\end{keyword}

\end{frontmatter}{}

\section{Introduction\label{sec:Introduction} }

The novel life threatening virus `Severe Acute Respiratory Syndrome
Coronavirus 2 (SARS-CoV-2)', responsible for Coronavirus Disease -
2019 (COVID-19), is one of the greatest challenges that the world
is facing today. This is a respiratory virus and can spread through
human to human contact. Therefore, to prevent infection and to slow
transmission of COVID-19, the following physical distancing guidelines
are specified by WHO \citep{who2020coronavirus}:
\begin{itemize}
\item Maintain at least 1 metre distance between you and people coughing
or sneezing. 
\item Stay home if you feel unwell. 
\item Practice physical distancing by avoiding unnecessary travel and staying
away from large groups of people. 
\end{itemize}
With millions of people infected worldwide and thousands dead, the
Covid-19 pandemic is showing no sign of abating. A vaccine is yet
to be found, and physical distancing through lockdowns seems to be
the only way to slow the spread. However, the lockdowns are also pushing
the world economy to the brink \citep{gita2020lockdown}. Millions,
in countries across the world are facing the bleak reality of job
losses, business closure and economic hardship. Hence, it has become
imperative to find ways to allow a section of the society - those
who have either recovered or are unexposed to the disease, carry on
with their day to day business operations to kickstart the economy.
What is desired is a system, that will enable easing of the lockdowns,
allowing countries to continue with their economic activities, perhaps,
in a reduced capacity, while still strictly maintaining appropriate
physical distancing norms for such relaxations. Consequently, strategies
have to be put in place that will assist people in practising the
distancing norms in public places. Modern communication technologies
and the ubiquity of smartphones use can be leveraged to devise such
mechanisms.

Many countries have developed mobile applications to combat the spread
of the virus by adopting technologies such as bluetooth and GPS data
for contact tracing, which is a mechanism for tracing people who had
come in close contact with some one who is infected with the virus
\citep{eames2003contacttracing}. Such mobile applications collect
information about persons who have been in close proximity with an
individual from their smart phones. If the individual is ever diagnosed
with the virus, every person who had possibly been near that infected
individual, during the period in which he was contagious, can be traced
and directed to take measures such as self-quarantine and or testing
\citep{normile2020coronavirus}. However, such applications facilitate
identification of only those people who have already been exposed
to an infected individual. While there is no doubt that contact tracing
can contribute immensely in containing the virus, we have certainly
reached a stage where economic activities also have to be allowed
in a disciplined manner. Therefore, it is highly desirable that systems
be put in place that can aid in prevention of exposing an individual
from infection while s/he is engaged in her/his day to day business
activities. Applications should be devised in such a way that it's
users can be given prior intimation about the Covid status of people
in their vicinity or of people that they plan to meet. There should
be provisions in the application that alerts users before they are
entering a zone or an area that is certified to be a covid hotspot
zone. In such a scenario, the person receiving such alerts can then
take preventive measures. Since the success of such applications depend
largely on people participation, user privacy concerns should be taken
into account and there should be inbuilt mechanisms to provide incentives
to people for their participation. The data accumulated through such
a public initiative should be available to general public for scrutiny
and use, so that they can reap it's benefit in their individual businesses. 

\subsection{Our Contribution \label{subsec:OurContribution} }

In this paper, we propose a framework that can be used by people to
protect themselves from infections while they are involved in their
activities. Individuals can use their smart phones as a digital-pass
to convince authorities, institutes and business establishments that
s/he is safe from the virus and has not come in close contact with
anyone who is infected by the virus. This will allow him or her to
move freely without restrictions to accomplish their tasks and allow
for some semblance of a normal life. For individuals who fail to do
so, it will be implied that he is infected or is under quarantine
and is not supposed to roam around in public places. Therefore, access
to public/private facilities will be denied to such individuals till
his/her quarantine period is over. 

We propose the use of blockchain technology for secure record keeping of persons using the application.
The covid-19 status of individuals along with other relevant information
such as age, pre-existing medical conditions etc., can be entered into
the blockchain, while ensuring privacy and anonymity to the individuals.
This public ledger will be made available to all the stakeholders.
Given the immutable nature of the blockchain, the ledger will be available
on a read only basis to these stakeholders. The stakeholders will
include entities such as the offices and health centers of the state,
 central or federal government, medical research and development centers, private hospitals,
business houses, and other organizations. 

\subsection{Paper Organization}

\label{subsec:PaperOrganization} The rest of the paper is organized
as follows. In section \ref{sec:SimilarWork}, similar work carried
out recently are discussed. In section \ref{sec:OverviewOfBlockchain},
we present a brief overview of the blockchain technology. In section
\ref{sec:ProposedFramework}, the proposed blockchain based framework
is presented. In section \ref{sec:discussion}, we discuss the proposed
work with reference to existing applications and platforms that were
developed recently. Finally, in section \ref{sec:Conclusion}, we
conclude the paper.

\section{Similar Work}

\label{sec:SimilarWork} Countries around the world have adopted mobile
technologies to mitigate the spread of corona virus and are also relying
on such technologies to provide data to augment their decisions making
on the exit strategy for lockdowns~\citep{coronastrategy}. China
\citep{chinaapp}, Singapore \citep{singaporeapp}, South Korea, Israel~\citep{israelapp}
and Australia~\citep{covidsafeaustralia} have taken the lead in
asking their citizens to install surveillance apps to facilitate contact
tracing. India has developed Aarogya Setu app for the same purpose
\citep{aarogyasetu}. These government initiatives have little or
no privacy protections built into their systems which have raised
concerns of the applications being used for surveillance beyond their
stated purpose. For example, the South Korean app uses GPS location
data and citizens are required to provide their real names and their
government issued identity numbers \citep{coronatrackingabc}. This
means that privacy is non-existent in the contact tracing mechanism
and citizens using the app can be tracked anytime and anywhere. Singapore's
\emph{TraceTogether }app stores anonymised IDs of nearby phones exchanged
through bluetooth, in encrypted form . The IDs are generated by encrypting
with a private key held by the Ministry of Health and hence can be
decrypted by only this government agency. This does not reveal the
users identity to other parties but is known to the government. Hence,
users have no privacy from government surveillance. India's Aarogya
Setu seeks the users bluetooth connectivity and location data at the
time of installation, which is sent to government servers effectively
allowing government agencies to know the users whereabouts. Hamagen
( Hebrew for the shield ) app, endorsed by Israel's Ministry of Health,
periodically (typically one hour) downloads anonymous GPS location
data of patients diagnosed as covid positive from the MoH cloud service.
It then compares this list with location data Hamagen has stored of
the user in the phone. This cross-referencing can then determine if
the user has been in presence of anyone who has been tested as positive.
User location is not sent to the government servers and the comparing
of location data is done on-device of the user. Location history and
WIFI networks that the user came within range, is stored on device for two-weeks.
Future versions would keep track of bluetooth connections and
data over sound. The Israeli government has plans of making the app
open source in the near future.

Private organisations have also jumped in the fray and are offering
mobile app based solutions. PHBC~\citep{phbc} is a consortium of
various health stakeholders such as universities, health care providers,
government agencies, etc which has developed virus record keeping
blockchain that allows the monitoring and verification of workplaces
and geographical zones that are free from corona virus~\citep{virusblockchain}.
Such zones are designated as safe zones. The system tracks the whereabouts
of uninfected persons and aims to constrain their entry into uninfected
areas if they have visited infected areas by requiring them to be
in the quarantine zone before they are allowed into the safe zone.
The system integrates AI and GIS technologies and draws information
in real time from agencies that provide up to date virus infection
information.

IT Researchers have also proposed a variety of approaches to control
and track the spread of the virus. The authors in~\citep{nguyen2020blockchain}
propose a blockchain and AI based 4 layer framework where corona virus
data is collected from various sources such as laboratories, hospitals,
social media, patient generated data and wireless network operators.
They propose to ensure privacy of such data by use of the blockchain.
AI models can then harness the data to provide solutions for outbreak
estimation, virus detection, analytics, assisting in vaccine development
and predicting of similar virus outbreak in future. However, the framework
is conceptual and does not provide implementation details.

Covid-watch is group of volunteers spread over multiple continents
and countries comprising security, policy and public health experts.
The group has developed a bluetooth based privacy-preserving mobile
app that aims to reduce the spread of corona virus~\citep{covidwatch}.
Their scheme works by generating and broadcasting random numbers whenever
a user of the app is in close proximity to another user. Bluetooth
signal strength is used to estimate proximity. In every phone a record
is kept of each random number the device has transmitted or received.
If any phone app user is diagnosed as positive the local health authority
gives them a permission number which after verification is sent to
a public server along with the persons history of sent and received
random numbers and is also transmitted to all other phones. If any
of the phones find a match with its stored list of random numbers,
it means that they were in close contact with the infected person.
While simple and elegant, the scheme can be compromised if there is
a man-in-the-middle attack on the bluetooth connections. Also, this
app is designed only for contact tracing. However, a lot more data
could be collected that might be of great use for epidemiological
purposes which the app is not designed for. 

De Carli et al. presents a privacy-preserving mobile tracing application
- WeTrace in~\citep{DeCarli2020}. Mobile phones that have the application
installed, periodically broadcast the app generated public key. Like
Covid-watch, other phones in close proximity ( estimated using Bluetooth
Low Energy) record these public keys locally. When users change their
status from ``not infected'' to ``infected'', messages encrypted
with all the public keys available in its local storage is sent to
the backend, which broadcasts it to all users. Those who were in close
proximity to this user can now decrypt the message with their private
key and know they were near an infected person. The backend does not
store any data, it only serves to broadcast messages to the mobiles
in the system. However as the authors acknowledge, such a system can
suffer from DDOS attacks on the backend, the backend can be susceptible
to impersonation and rouge users may issue false notifications.

Torky and Hassanein propose a blockchain based framework for mitigating
the spread of coronavirus~\citep{Torky2020}. Their framework consists
of four subsystems - Infection Verifier Subsystem (IVS), P2P Mobile App, Blockchain
Platform and Mass Surveillance System. The IVS is the part of the
system that records covid positive persons in the blockchain by using\emph{``infection
patterns''} which are regular expressions. The patterns derived from
this regular expression, the \emph{``infection instances''} are
used to digitally represent people or places who have been infected
by this patient. These are also recorded in the blockchain. A finite
automaton can verify if a regular expression instance conforms to
the infected pattern which will indicate a high probability of contagion.
By representing people as regular expressions, the system does anonymise
the users. These instances are also used in their proposed P2P mobile
application. In their scheme, the authors use the blockchain for storing
the infected pattern and all confirmed cases (infected instances)
based on the pattern. The P2P application serves to notify the populace
that they might have been in contact with an infected person or been
to a place visited by an infected person. The Mass Surveillance System
component is responsible for contact tracing and for identifying the
public places the infected person may have visited in the last few
days, which is sent to the blockchain to be stored as infected patterns
and infected instances. However, it is unclear what underlying communication
technology is employed to implement this mass surveillance scheme.
It is also unclear who the block miners are in this system and what
incentives they have. In Table \ref{tab:ComparisonWithOtherApps},
a summary of important features of the various schemes discussed in
this section is presented.

\section{Overview of Blockchain}

\begin{figure}[H]
\centering{}\includegraphics[width=0.7\columnwidth]{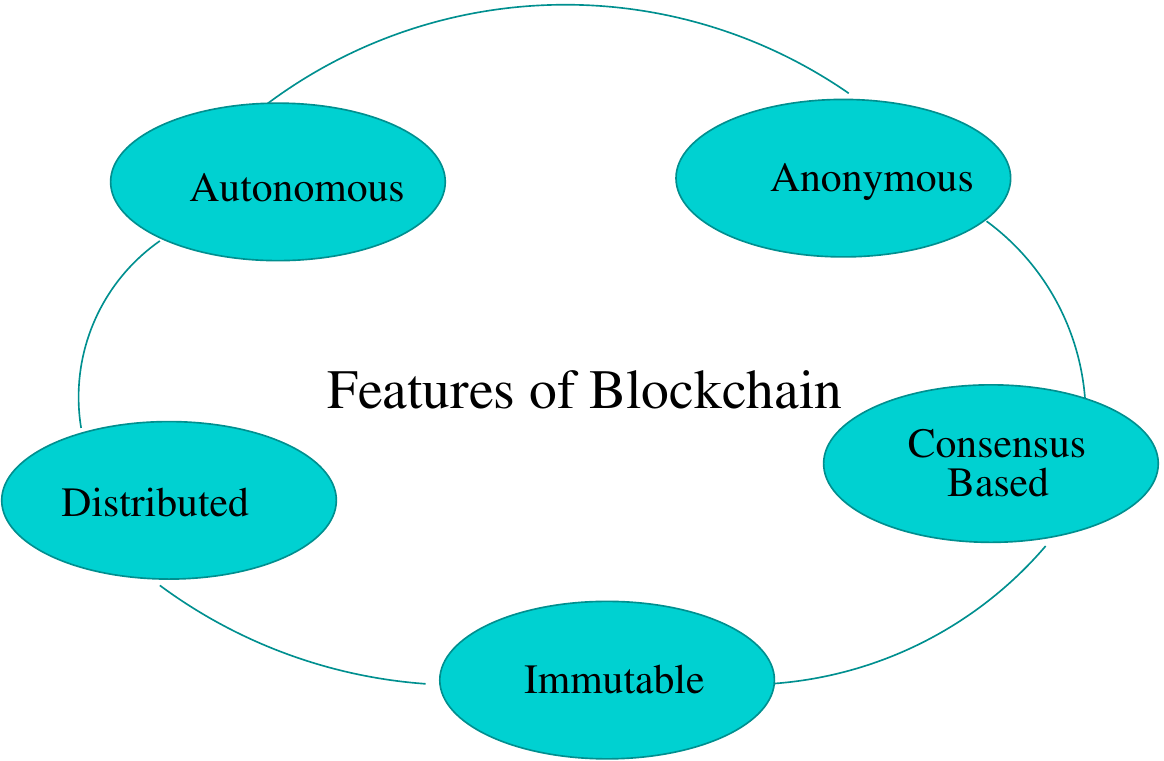}
\caption{Features of blockchain.}
\label{fig:Features-of-Blockchain}
\end{figure}

\label{sec:OverviewOfBlockchain} Blockchain as a concept was first
coined in the white paper of Bitcoin by Satoshi Nakamoto in 2009~\citep{nakamoto2019bitcoin}.
It is a mechanism for the storage of decentralized, time stamped and
immutable data. As it is decentralized and distributed in nature,
it is not owned by a single authority and is kept as a distributed
ledger. As shown in Figure~\ref{fig:Features-of-Blockchain}, some
intriguing features of blockchain that motivated us in adapting this
technology in our proposal, are described as follows:\citep{feng2019survey}

\paragraph*{Autonomous}

Transactions in a blockchain can be carried out without the involvement
or control of a central authority or a single authority governing
the network. This architecture provides the benefits of a decentralized
system with reduced operational costs and avoiding performance bottlenecks
at the central location. Any node can add transactions and review
them at any time.

\paragraph*{Distributed}

Transactions in a blockchain are virtually impossible to tamper as
they are recorded in blocks spread over the entire network with each
block validated by other nodes. In such a scenario any false transaction
or changes can be detected very easily.

\paragraph*{Immutable}

All transactions and blocks added chronologically to a blockchain
ledger are verifiable and all changes are traceable. Additionally,
a consensus mechanism synchronizes the blocks in all the nodes. These
measure ensures that data in the blockchain cannot be altered and
changed.

\paragraph*{Consensus based}

New transactions and blocks are added in the blockchain based on a
previously agreed upon mechanism called consensus method. All the
stakeholders update their copy of the ledger by adding the same block
to their individual blockchains (ledger) with the help of this consensus
mechanism. Therefore, each individual at a particular instance of
time has the same copy of the blockchain, synchronized with every
other individual.

\paragraph*{Anonymous}

In blockchain, information about individuals can be recorded with
a generated address, which does not reveal the real identity of the
individual. Thereby, facilitating anonymous storage of individual
records.

\begin{figure}[H]
\centering{}\includegraphics[width=0.7\columnwidth]{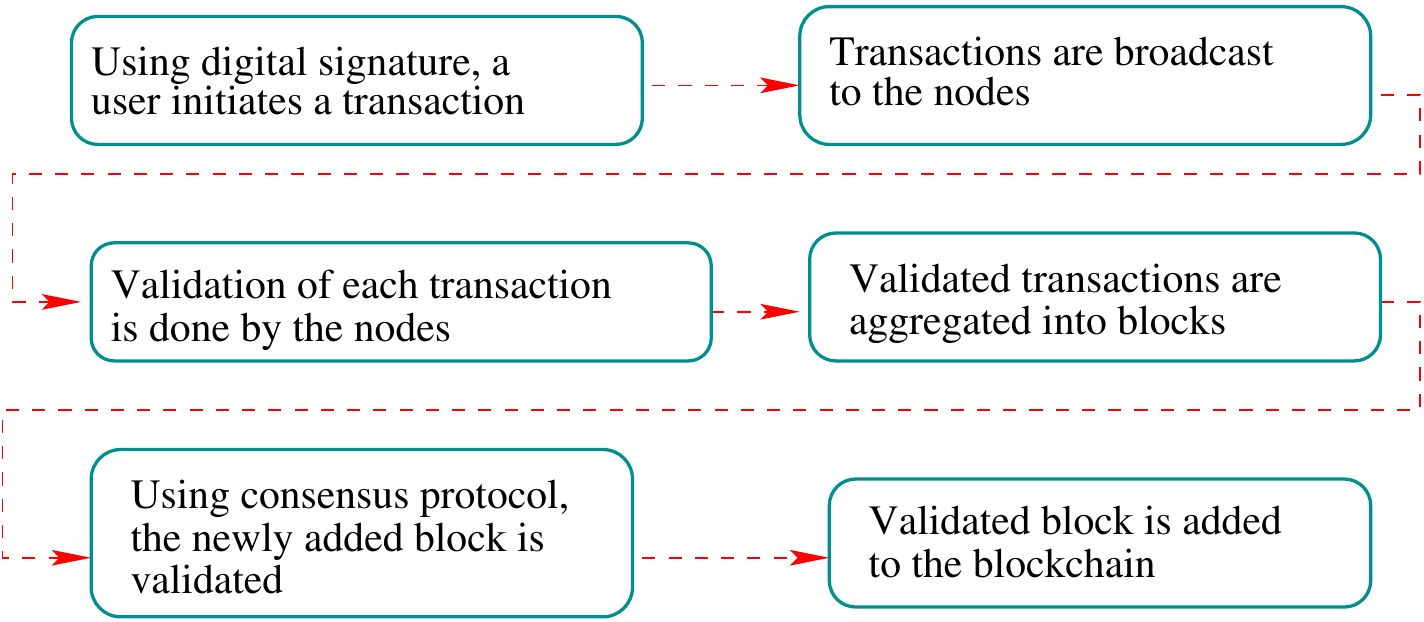}
\caption{Stages of a blockchain.}
\label{fig:BlockChain}
\end{figure}

A schematic view of how a Blockchain works is shown in figure~\ref{fig:BlockChain}
.
 When a new transaction is initiated, it is broadcast among the nodes,
and based on a consensus algorithm, a miner adds the transaction into
the block. The \emph{nodes} in Blockchain can be any device (computer,
mobile phone) that contains a copy of the Blockchain. The \emph{miners}
are nodes that create blocks after validating new transactions and
adhering to a accepted consensus mechanism. Without involvement and
authentication from any central authority, these blocks are validated
using the \emph{consensus algorithm} which establishes mutual trust
and enables a decentralized network to take a decision. Consensus
algorithm plays a significant role in maintaining the safety and regulation
of the Blockchain. Common consensus algorithms are `proof of work',
`proof of stake' and `proof of existence'~\citep{sankar2017survey}.

\emph{Proof of Work} (PoW) is a popular consensus method used by various
crypto-currencies. It's popularity is due to the fact that it provides
safety from various attacks. PoW is used to validate transactions
and add new blocks to the chain by solving a complex computational
problem. Miners race against each other to solve the problem and create
the blocks. The first miner to solve the problem gets to add the new
block into the blockchain and claim a reward. The reward is an incentive
mechanism that ensures the participation of miners in the blockchain.
PoW has some limitations like it is computationally intensive and
vulnerable to attack from 51\% of the nodes in the network. To overcome
the limitations of PoW method, \emph{Proof of Stake} (PoS) method
was introduced that determines the block creator based on a combination
of node’s wealth, the staking age and random selection. Selection
purely on the basis of the node's wealth would result in the node with
the most wealth having a constant advantage in block creation. Hence
a combination of various strategies have been adopted. The staking
age ensures that nodes that hold the stake longer have a better chance
to forge blocks which decreases the chances of malicious attacks.
As a reward transaction fees are given to the block validators.

\begin{figure}[H]
\centering{}\includegraphics[width=0.8\columnwidth]{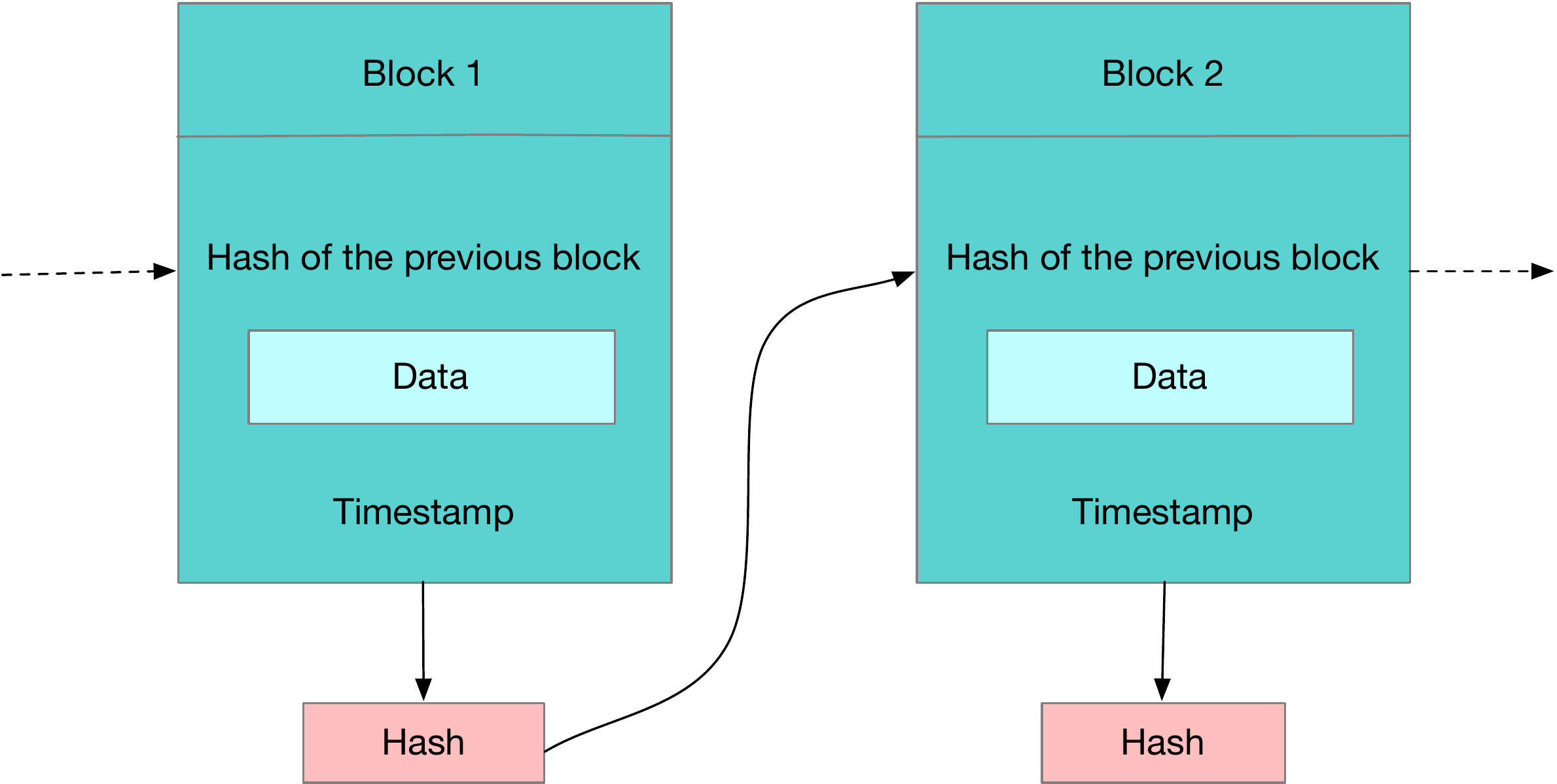}
\caption{Structure of a blockchain.}
\label{fig:Structure-of-Blocks}
\end{figure}

The Blockchain consists of \emph{chain of blocks} as shown in the
figure \ref{fig:Structure-of-Blocks}. 
Each block is arranged in chronological order, consisting of block
header and block body. Every block also contains the hash of the previous
block, its data, merkle root and timestamp. As each block contains
the hash value of the previous block, it is ensured that blocks are
linked and changes cannot be made in any one of the blocks without
being reflected in all subsequent blocks. This makes data in the
Blockchain virtually impossible to alter and change. A block contains
a number of transactions, and all these transactions have a hash associated
with it. Each pair of these hashes are then concatenated to compute
another hash and so on. Such hashing is used to build a tree of hash
values. The merkle root is then the hash of all the transactions a
block contains. \emph{Merkle Root} is used to provide integrity of
all the transactions in a block, as changing any one transaction will
change the merkle root.

Depending on various conditions, Blockchain can be categorized into
public, private and consortium.

\paragraph*{Public Blockchain}

Public Blockchains are fully decentralized, the ledgers are open and
transparent. Anyone can be part of this Blockchain as a miner or as
a node without any restriction. Example of public Blockchain are Ethereum,
Monero, Bitcoin, etc. Bitcoin is a decentralized cryptocurrency in
the absence of a central administrator in which everyone can invest.

\paragraph*{Private Blockchain}

Private Blockchains are similar like that of public Blockchain following
a set of conditions.It is governed by an organization and only
those who have permission or invitation to access can participate.
Private Blockchains are also called permissioned Blockchain. It is
also a decentralized peer-to --peer network but the participants
are decided by an organization. Hyperledger is an example of Private
Blockchain.

\paragraph*{Consortium Blockchain}

In a Consortium Blockchain, a group of companies or stakeholders comes together to develop
a Blockchain. It is partly private and partly public Blockchain
and is also referred to as a semi-decentralised Blockchain. Consortium
Blockchain consists of pre-determined group of nodes and validation
of blocks are also determined by specific group. Also in this type
of Blockchain, certain operations may be open to other participants. Example of
of consortium Blockchain are R3, EWF, Libra, etc.

Blockchain is widely used in the field of cryptocurrency - Bitcoin
being its most popular implementation. However, other than Bitcoin,
Ethereum, Monero, Litecoin are some examples of cryptocurrencies where
blockchain is used to maintain the distributed ledger of transactions.
Furthermore, Blockchain technology has found use in areas other than
cryptocurrency and is transforming these fields. A few areas where
blockchain is proving particularly effective are: 
\begin{itemize}
\item \emph{Electronic Health Records (EHRs)} based on Blockchain provides
patients access to immutable and complete records of their own medical
history without any service providers, as opposed to the traditional
method where electronic records of patient's health are held by the
respective hospitals. \citep{guo2018secure} 
\item \emph{Product Ownership Management System (POMS)}: Blockchain is used
for anti-counterfeits in the RFID enabled supply chain in which a
customer can reject purchase of counterfeits products even with a
genuine RFID tag if the seller does not have the ownership of the
product. The ``proof of possession balance'' of public ledger of
the blockchain is used to verify th said ownership. \citep{toyoda2017novel} 
\item In \emph{cloud computing}, metadata that keeps a record of the history
of creation and operations that have been performed on a cloud data
object is of utmost importance. It is used to provide accountability,
forensics and privacy. Blockchain-based data provenance architecture
can be used to ensure transparency of such data, and can provide tamper-proof
records which is essential for transparency of data accountability
and can enhance availability and privacy of provenance data. \citep{liang2017provchain} 
\item A Blockchain-based access control model in IoT is introduced to manage
access control, where smart contract is used for contextual access
control policies to make authorization decisions. This blockchain
based solution ensures user privacy by avoiding the use of a third
party to handle and implement access control policies. \citep{ouaddah2017towards} 
\item A Blockchain-enabled platform can be used for processing insurance
transactions to speed up the processes in the insurance industry,
to make client data confidential, to reduce operational costs, transaction
processing times and payment settlement duration and provide security
in the whole insurance mechanism.\citep{raikwar2018blockchain} 
\end{itemize}
Given the inherent benefits of blockchain technology, it was felt
that it can contribute immensely in designing a platform that can
be used in fight against the spread of the COVID-19 virus.

\section{The Proposed Blockchain Framework}

\begin{figure}[H]
\centering{}\includegraphics[scale=0.4]{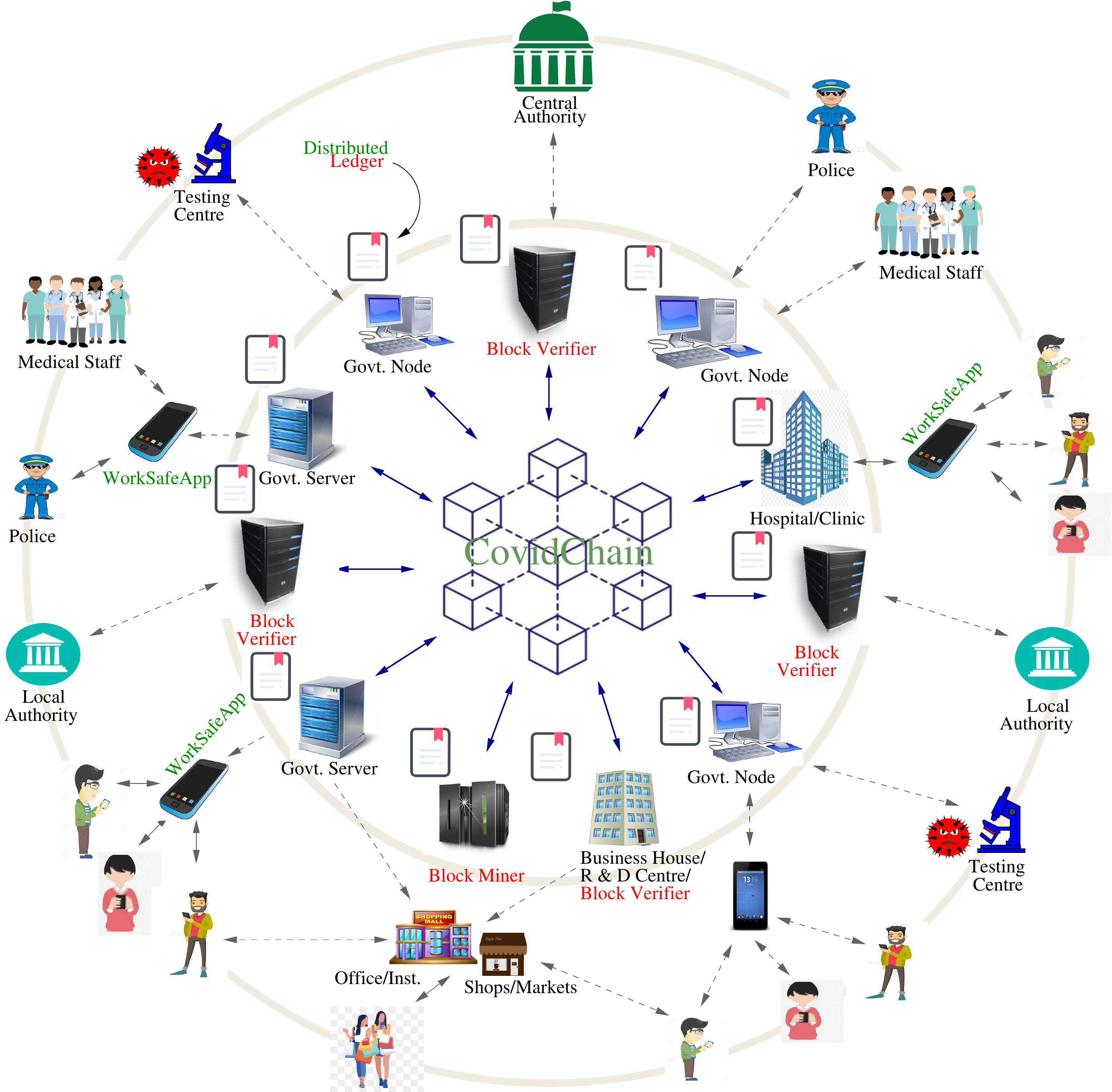} \caption{Blockchain framework for protection against COVID-19.}
\label{fig:BlockchainFramework}
\end{figure}

\label{sec:ProposedFramework} In this section, we present our blockchain
based framework called `CovidChain' for protection against COVID-19.
Compared to majority of the applications that have been developed
in the recent past with an intention to contain the spread of COVID-19,
our proposal differs with regards to the following aspects. 
\begin{itemize}
\item Most of the apps are specifically designed for tracing and alerting
those people who might have come in contact of an infected person.
Such a mechanism helps the intimated people to take precautions with
regards to their own health and social behaviour. However in addition
to this aspect, in our framework, we focus on another issue: which
is that of allowing those individuals of the society who have not
tested positive or have not come in close contact with a corona positive
person, to continue with his/her economic activities while protecting
himself/herself against the virus. 
\item The data generated in most of the apps are stored in a single location
that is under the control of a single authority. In our proposal,
the generated data is recorded in a consortium blockchain in the form of a distributed
ledger that is accessible to every section of the society. Such an
approach will increase transparency and will allow the entire population
to garner it's benefit. It also helps in avoiding confusion about
the current status of the infection as reliable data is available
to everyone at the same time. 
\item The framework is designed keeping privacy, anonymity and incentive
for the stakeholders in mind. 
\end{itemize}
The various stakeholders and entities associated with the framework
are as follows Figure \ref{fig:BlockchainFramework}. 
\begin{itemize}
\item \textit{Medical professionals}: They are people who perform random
testing. In addition to their testing kits, they are equipped with
a bluetooth enabled smartphone and Internet connectivity. They have
read-only access to the ledger and generates transactions comprising
of covid status of individuals based on their test report (for example:
+ive, -ive, number of days of quarantine required, etc.). 
\item \textit{Hospitals/Clinics/Laboratory}: They have read-only access
to the ledger and can generate transactions comprising of covid status
of individuals based on their test report (for example: +ive, -ive,
number of days of quarantine required, etc.), for onward transmission
to the block miner. 
\item \textit{Testing Centres}: These are centres established by the Govt.
with facility for COVID-19 testing. They are equipped with a bluetooth
enabled smart phone, computer and Internet connectivity. They have
read-only access to the ledger and generates transactions comprising
of covid status of individuals based on their test report (for example:
+ive, -ive, number of days of quarantine required, etc.). 
\item \textit{Govt. nodes}: These are computers established by the authority
having read-only access to the ledger. They facilitate testing centres,
law enforcement personals, medical professionals and individuals to
access the ledger and communicate transactions generated by them. 
\item \textit{Business houses, Research and Development Centers, etc.}: They are nodes with
read-only access to the blockchain. They are interested in the data
getting appended in the blockchain for research and business purpose.
Such organizations can even participate in the system as a Block Evaluator. 
\item \textit{Institutes, shops, offices, etc.}: These establishments have
read-only access to the ledger through a Govt. node or a Business
house. They are interested in the data that is getting generated,
so that they can use it for smooth running of their business activities. 
\item \textit{Individuals}: Through their smart phones, they have read-only
access to the ledger through a Govt. node or a node belonging to a
business house. They can use the data to protect themselves from infection
as they go about their day to day business activities. 
\item \textit{Law enforcement personals}: These are people like policeman
etc., who suggests quarantine to individuals with recent travel history
or having been contact traced to covid positive individuals, by generating
such transactions for onward transmission to the block miner. They
also demarcate a locality, district, etc. into red, orange, green
zone based on number of people infected in a locality, so that people
outside such zones can be alerted about their proximity to such a
zones. Such personals have to be equipped with Internet and bluetooth
enabled smartphones or computers. 
\item \textit{Block Miner}: It is a centrally located node in the blockchain
that receives all the transactions generated by the various stakeholders.
These transactions are then put together into blocks. 
\item \textit{Block Verifier/Validator}: These are nodes that validates all the transactions by examining their digital signatures. A Block Validator is either owned by the state/central government of a country or are owned/monitored by establishments that are nominated by the civil society. 
\item \textit{Local Authority}: A local government of a state or a province. 
\item \textit{Central Authority}: The central government of a country. 
\item \textit{WorkSafeApp}: It is a smart phone application that is used
as an interface to interact with CovidChain. WorkSafeApp requires
the smart phone of an individual to be bluetooth enabled. Like many
recently developed applications for contact tracing through bluetooth
\citep{aarogyasetu}\citep{covidwatch}\citep{singaporeapp}, WorkSafeApp
also collects anonymized tokens from the smart phones of all such
individuals with whom a person has come in close proximity in the
past few days. If the person is ever tested positive for COVID-19,
all the collected tokens are mapped to their corresponding contact
numbers and all such individuals are either tested for COVID-19 or
are suggested to be in quarantine by medical professionals or law
enforcement agencies. WorkSafeApp also has a self assessment module,
using which a user can assess his health status by answering certain
queries. Depending on the answers and severity of the symptoms, the
module suggests people to either self quarantine or to get in touch
with a Testing Centre. As the blockchain data is publicly available,
applications similar to WorkSafeApp with CovidChain as the backend
data may be developed by individual businesses. 
\end{itemize}
We split the functionality of the entire framework, including the process of adding a block
in the blockchain, into the following activities. 

\subsection{App Initialization}

When WorkSafeApp is installed in a smart phone, a pair of public-private
key is generated by taking an OTP verified phone number as input from
the user. Hence forth, the phone number is neither recorded nor used
for any other purpose. For all subsequent purposes, the public key
is used as it's identity by the user.

\subsection{Transaction Generation}

A transaction may be of two types, viz., transactions indicating individual
status `$T_{Ind}$' and transactions indicating location status `$T_{Loc}$'.
`$T_{Ind}$' is a statement that represents the covid status of an
individual as determined by a medical professional, a testing centre,
a hospital or a law enforcing personal. It has three components, viz.,
the hash of the public key of the individual `$H(K_{pub_{Ind}})$',
the COVID status `$CS$' of the user, the $Date$, the $Time$ and
a minimal set `$S_{Epid}$' of epidemiological information like age,
gender, blood group, state/province, pre-existing disease, etc., that
is required by the Govt. to make a general study on the overall data,
so that new knowledge about the proliferation pattern of the virus
may be derived. `$S_{Epid}$' should contain minimum number of attributes
required for data analysis. It should not contain information like
House-no, Street-no, etc. that may compromise anonymity and privacy
of an individual. Before including the set `$S_{Epid}$' in the transaction,
it is encrypted using the public key `$K_{pub_{CA}}$' of the Central
Authority. A transaction is uniquely identified by a transaction identity
`$TID$', which is the hash value of the transaction. This is further
encrypted with the Testing Centre's private key `$E_{K_{pri_{TC}}}$',
for generating a digital signature `$DS_{TC}$' so that the genuineness
of the transaction may be verified by the Block Validators.
\begin{table*}[t]
\noindent \centering{}\caption{Acronyms and symbols\label{tab:Acronyms}}
\begin{tabular}{|c|l|}
\hline 
\textbf{{Acronym/Symbol}} & \textbf{{Description}}\tabularnewline
\hline 
$T_{Ind}$ & Transaction related to CS of an individual.\tabularnewline
\hline 
$T_{Loc}$ & Transaction related to CS of a zone.\tabularnewline
\hline 
$K_{pub_{Ind}}$ & Public key of an individual.\tabularnewline
\hline 
CS & COVID-19 (C19) status.\tabularnewline
\hline 
$S_{Epid}$ & Epidemiological information about an individual.\tabularnewline
\hline 
$S_{Enc}$ & Encrypted Epidemiological information.\tabularnewline
\hline 
$K_{pub_{CA}}$ & Public key of Central Authority.\tabularnewline
\hline 
$TID$ & Transaction identity.\tabularnewline
\hline 
$E_{K_{pri_{TC}}}$ & Private key of a testing centre.\tabularnewline
\hline 
$DS_{TC}$ & Digital signature of testing centre.\tabularnewline
\hline 
$H$ & One way hash function.\tabularnewline
\hline 
$E$ & Encryption function.\tabularnewline
\hline 
$D$ & Decryption function.\tabularnewline
\hline 
$lt$, $ln$ & Latitude, longitude.\tabularnewline
\hline 
$radius$ & Radius of a zone.\tabularnewline
\hline 
$zoneType$ & Zone type: red/orange/green.\tabularnewline
\hline 
$zoneId$ & Zone identity.\tabularnewline
\hline 
$DS_{LEA}$ & Digital signature of Law Enforcing Agency.\tabularnewline
\hline 
$K_{pri_{LEA}}$ & Private key of Law Enforcing Agency.\tabularnewline
\hline 
\end{tabular}
\end{table*}

Therefore, 
\begin{equation}
S_{Epid}=\{age,gender,state,..\}
\end{equation}
\begin{equation}
S_{Enc}=E_{K_{pub_{CA}}}(S_{Epid})
\end{equation}
\begin{equation}
TID=H(H(K_{pub_{Ind}}),CS,Date,Time,S_{EncEpid})
\end{equation}
\begin{equation}
DS_{TC}=E_{K_{pri_{TC}}}(TID)
\end{equation}
\begin{equation}
T_{Ind}=\{TID,H(K_{pub_{Ind}}),CS,Date,
Time,S_{Enc},DS_{TC}\}
\end{equation}
Where, `$E$' is an encryption function and `H' is a one way hash
function like SHA-256. The various symbols and acronyms used in this
section are presented in Table \ref{tab:Acronyms} for quick reference.
$T_{Loc}$ is a transaction made by a local authority or a law enforcement
agency that identifies a locality, district, etc., as a red zone ,
orange zone or a green zone, based on the number of infected people
in a particular area, so that people outside such zones can be alerted
about their proximity to such zones. Such transactions help the authority
to disseminate (and people to acquire), timely and reliable information about crucial decisions. To demarcate a locality,
the GPS coordinates (`lt', `ln'), radius of the zone `r', date, time,
zone type `$zoneType$' (red/orange/green) are recorded in a transaction.
Zone identity `$zoneId$' which is the hash of the location coordinate,
`TID' which is the hash of the entire transaction, and the digital
signature of the authority `$DS_{LEA}$' that is obtained by encrypting
the `TID' with the private key of the Law Enforcing Authority `$K_{pri_{LEA}}$',
are also recorded in a location transaction. Therefore, 
\begin{equation}
zoneId=H(lt,ln)
\end{equation}
\begin{equation}
TID=H(zoneId,lt,ln,r,
Date,Time,zoneType)
\end{equation}
\begin{equation}
DS_{LEA}=E_{K_{pri_{LEA}}}(TID)
\end{equation}
\begin{equation}
T_{Loc}=\{TID,zoneId,lt,ln,r,Date,Time,
zoneType,DS_{LEA}\}
\end{equation}
After a transaction is generated ($T_{Ind}$ or $T_{Loc}$), it is
transmitted to the Block Miner. 

\begin{figure}[H]
\centering{}\includegraphics[width=0.6\columnwidth]{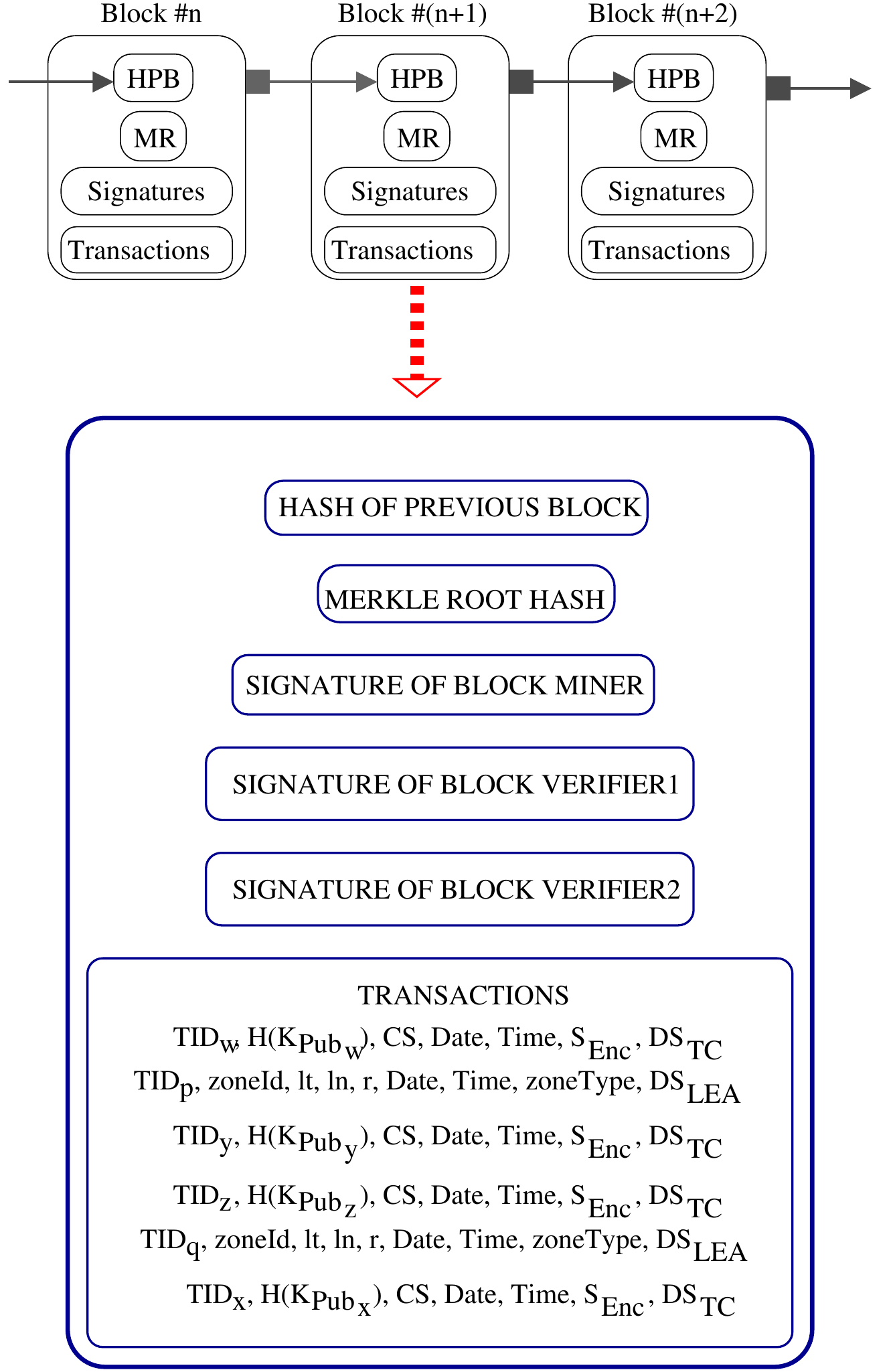} \caption{Structure of a block in CovidChain.}
\label{fig:BlockStructure}
\end{figure}

\subsection{Block Mining}

\label{subsec:BlockMining} 
\begin{figure}[H]
\centering{}\includegraphics[width=0.65\columnwidth]{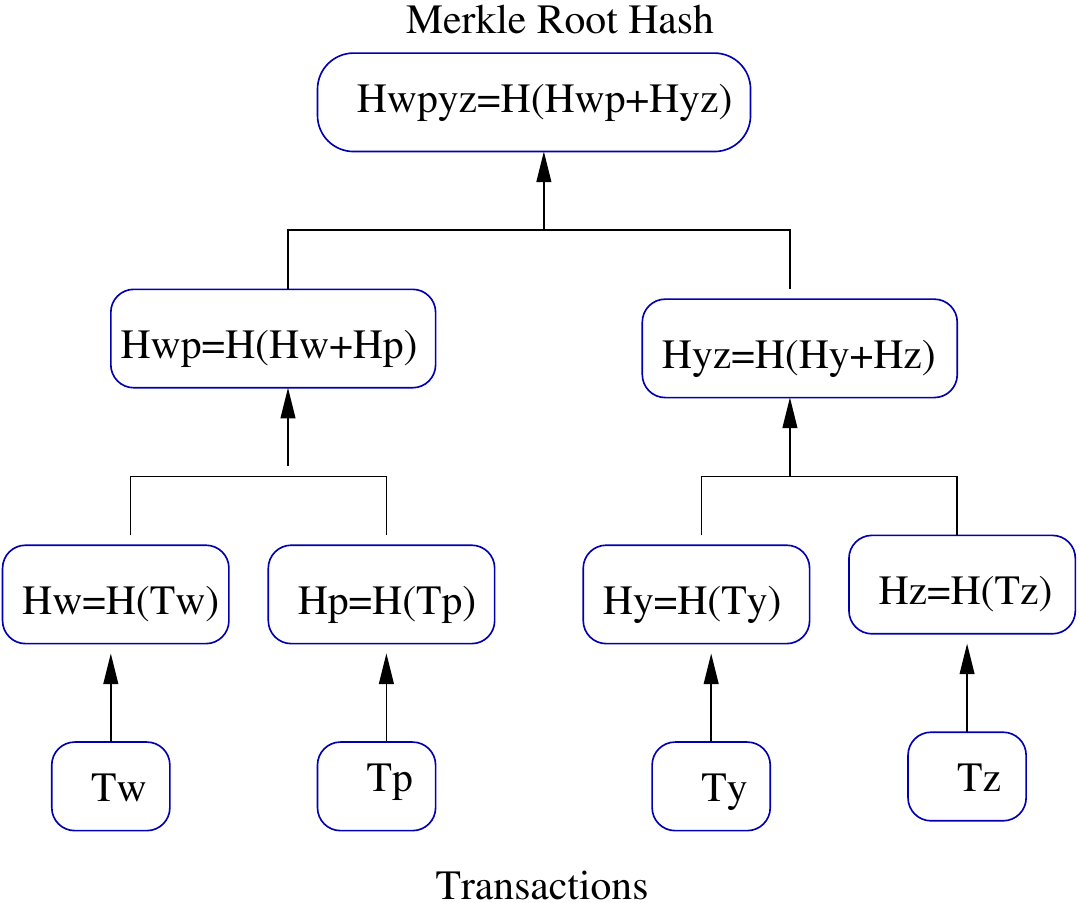}
\caption{Calculation of Merkle Root Hash from the Transactions.}
\label{fig:MerkleRootHash}
\end{figure}

The Block Miner contains a single or multiple
servers that receive the transactions from all the testing centres,
hospitals, etc. When the total size of the received transactions reaches
a predetermined block size (say 1 MB), they are bundled together into
a block (Figure \ref{fig:BlockStructure}). Each transaction in the
block is then verified by validating the digital signature of the
Testing Centres that signed it. The Merkle Root Hash (MRH) of the
block is calculated from the hash of all the transactions by pairing
the hash values and rehashing the sum of the pairs as depicted in
(Figure \ref{fig:MerkleRootHash}).
 After this, the Block Miner adds the MRH to the block and digitally
signs the MRH with its Private Key $K_{pri_{BM}}$. Therefore, if
`$T_{w}$', `$T_{w}$', `$T_{w}$' and `$T_{w}$' are the transactions
in a block, then the MRH `$H_{wpyz}$' of this block is calculated
and digitally signed as fols.

\begin{equation}
H_{w}=H(T_{w})
\end{equation}
\begin{equation}
H_{p}=H(T_{p})
\end{equation}
\begin{equation}
H_{y}=H(T_{y})
\end{equation}
\begin{equation}
H_{z}=H(T_{z})
\end{equation}
\begin{equation}
H_{wp}=H(H_{w}+H_{p})
\end{equation}
\begin{equation}
H_{yz}=H(H_{y}+H_{z})
\end{equation}
\begin{equation}
H_{wpyz}=H(H_{wp}+H_{yz})
\end{equation}
\begin{equation}
DS_{BM}=E_{K_{pri_{BM}}}(H_{wpyz})
\end{equation}
After this, the Block Miner forwards the block to a couple of Block
Verifiers/Validators `BV1' and `BV2', selected randomly from among all the available
Block Verifiers. Each transaction in the block is then verified by
validating the digital signature of the Testing Centres that signed
it; after which, the Block Verifiers digitally sign the MRH of the
block with their respective private keys $K_{pri_{BV1}}$ and $K_{pri_{BV2}}$.
\begin{equation}
DS_{BV1}=E_{K_{pri_{BV1}}}(H_{wpyz})
\end{equation}
\begin{equation}
DS_{BV2}=E_{K_{pri_{BV2}}}(H_{wpyz})
\end{equation}

The signed block is then sent back to the Block Miner. After receiving
the block, the Miner verifies the signatures of the Block Verifiers.
The block is then broadcast to all the nodes/stakeholders in the network.
Upon validation of the digital signatures, each node in the network
adds the block to their individual ledger (or their individual copy
of the blockchain). While adding the block to the blockchain, the
Hash of the Previous Block (HPB) is added to the current block.

\subsection{COVID status verification}

\begin{figure}[H]
\centering{}\includegraphics[width=0.85\columnwidth]{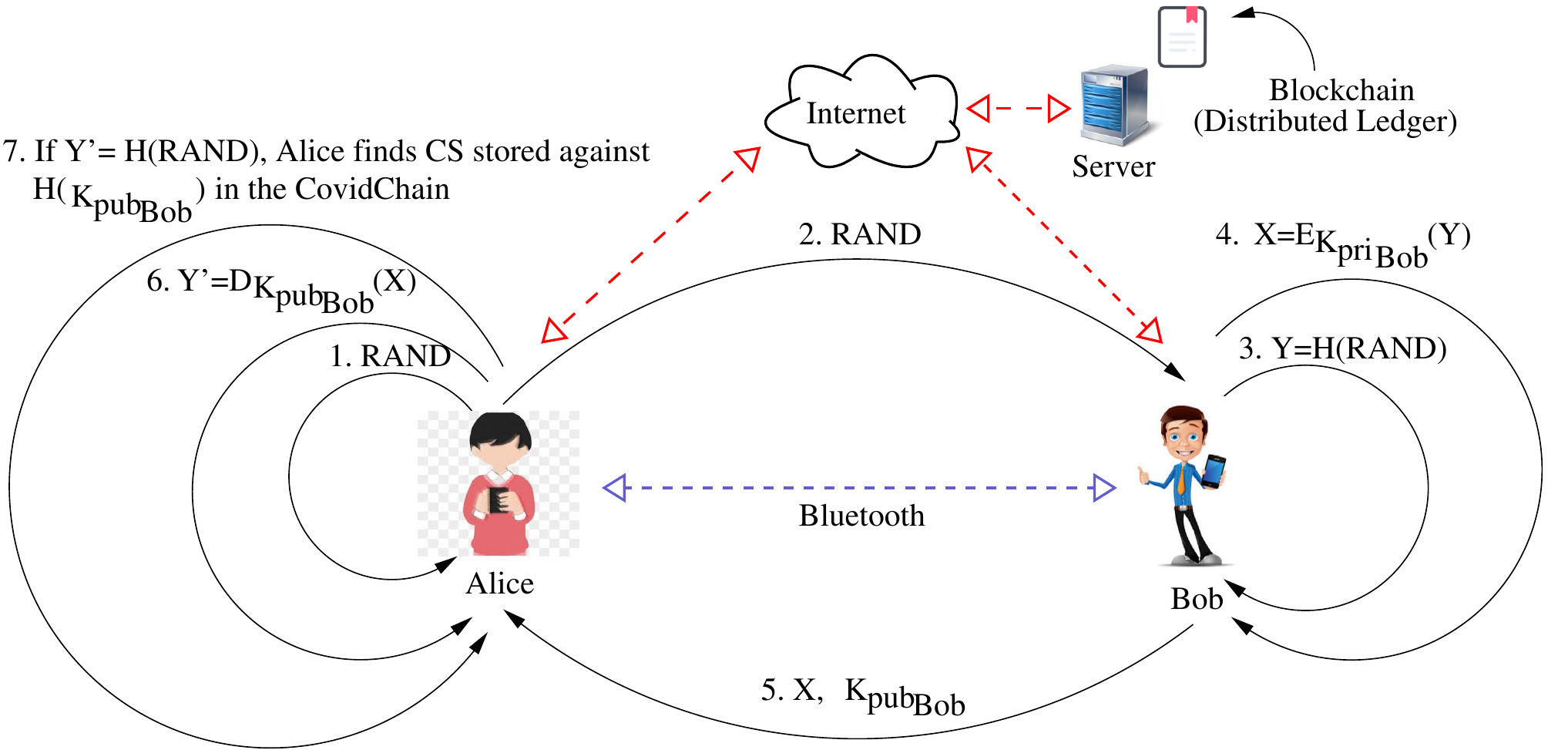}
\caption{COVID status verification.}
\label{fig:Verification}
\end{figure}

\label{subsec:CovidStatusVerification} `CovidChain', the proposed
blockchain framework will aid people by providing an additional layer
of protection from the virus while engaged in their activities. It
will enable entities in the society like offices, businesses, shops,
markets, educational institutes, traffic police, law enforcement agencies,
etc., to verify employees and clients at entry points based on data
available in the blockchain. An individual will also be able to verify
another individual through the bluetooth interface of his/her smart
phone as they engage in day to day business activity. For example,
Alice will be able to verify the covid status of Bob, a plumber, before
letting him into her premise. Similarly, Bob will be able to verify
the covid status of Alice before entering her premise. A delivery
boy will be able to verify his client's covid status through a web
interface, before setting off for delivery. Therefore, CovidChain
on one hand will allow Alice to verify the covid status of Bob and
on the other hand will allow Bob to present his smart phone as an
E-pass to Alice. The sequence of steps involved, when Bob presents
his credentials to Alice for verification, is follows (Figure \ref{fig:Verification}).
\begin{enumerate}
\item Alice: Generates a fresh random number $RAND$. 
\item Alice: Transmits $RAND$ and request for C19 status to Bob. 
\item Bob: Calculates the following 
\begin{equation}
Y=H(RAND)
\end{equation}
\begin{equation}
X=E_{K_{pri_{Bob}}}(Y)
\end{equation}
Where, $E_{K_{pri_{Bob}}}$ is the private key of Bob, `E' is an encryption
function and `H' is a one way hash function. 
\item Bob: Transmits `X', and his Public Key `$K_{pub_{Bob}}$' to Alice. 
\item Alice: Calculates the following. 
\begin{equation}
Y'=D_{K_{pub_{Bob}}}(X)
\end{equation}
Where, `D' is the decryption function. 
\item Alice: Compares $Y'$ and $H(RAND)$. If Alice finds that $Y'=H(RAND)$,
it is convinced that $K_{pub_{Bob}}$ is the public key of Bob. After
this, Alice finds the covid status of Bob from the CovidChain using
$K_{pub_{Bob}}$ (i.e., $H(K_{pub_{Bob}})$). 
\item Alice: If the COVID Status against $H(K_{pub_{Bob}})$ is neither
`+ive' nor `IQ' (ie., in quarantine), or if there is no transaction
against `$H(K_{pub_{Bob}})$', Alice may allow Bob to have access
to her premise with additional social distancing precautions. 
\end{enumerate}
Bob can also verify the covid status of Alice in a similar manner. 

\subsection{Accessibility of Data}

\label{subsec:Accssibility} With the rapid spread of the virus, the
numbers of infection, case fatality, etc., and administrative decisions
like declaring a zone to be a containment zone changes rapidly. In
such a fluid scenario, disseminating reliable timely information that
policy makers, epidemiologists, news agencies and people in general
depend heavily on, becomes a challenging affair. Especially so, when
the data is in control of a single authority. Since the proposed framework
is based on blockchain technology, the data is stored in distributed
ledgers that are open for public access. The data is immutable, updated
in real time, and is available for public scrutiny and audit. The
possibility of the data being manipulated by any stakeholder with
malicious intention is negligible because of its immutable nature.
The availability of the data to various sections of the society like
shops, businesses, offices, households, etc., will empower them to
take crucial decisions with regards to their economic activities.
For instance, allowing uninfected people or people who are not in
quarantine to join work, etc.

\subsection{Anonymity and Privacy of Data}

\label{sebsec:Anonymity} Anonymity and privacy of data related to
individuals is an important aspect for the success of such a platform.
If an individual gets the perception that his/her anonymity or privacy
of data is not being respected, s/he will be apprehensive about using
such an application. Such applications will be successful in its objective,
only if there is widespread adoption among the populace. Taking this
into cognizance, we have designed the framework on blockchain. In
our framework, any stakeholder including the Central Authority, having
access to the distributed ledger or the blockchain, will not be able
to link a particular transaction in the ledger with any individual,
because the anonymity of transactions are guaranteed as follows. 
\begin{itemize}
\item While creating a block, the Block Miner bundles the transactions in
the same order as it receives from the different stakeholders. Therefore,
a single block will have multiple transactions, originating at different
sources, shuffled together. This will make it extremely difficult
for anyone having access to the blockchain to link the data available
in a transaction to its origin. 
\item Since the Testing Centres, Hospitals, etc., records the Hash of the
Public Key in a transaction, any node/stakeholder having access to
the blockschain will not be able to traceback any data to any individual
because it will not be able to map a given Hashed Public Key to its
corresponding Public Key. 
\item When an individual (say Alice) has to present his identity to another
individual (say Bob) located a few feet away, s/he transmits his/her
Public Key $K_{pub_{Alice}}$ through the bluetooth interface. Bob
searches through the ledger for any transaction that is recorded against
$H(K_{pub_{Alice}})$. If Bob finds any such transaction, the only thing
he gets to know about Alice is her COVID status and nothing else,
because the rest of the epidemiological data is encrypted with the
private key of the Central Authority. 
\item The Central Authority, having access to the epidemiological data through
it's private key, can only link the data in a transaction to a hash
value that cannot be mapped to its corresponding Public Key. Therefore,
it cannot link any data with any particular individual. Such data,
will only help the authority to carry out epidemiological analysis
without compromising anyone's privacy. 
\item While analyzing the epidemiological data, the Central Authority cannot
traceback a particular transaction of interest to its origin. But,
if the Central Authority wishes, it can collect the public key of
an individual through the bluetooth interface (say at a entry point
to a public facility) and locate his epidemiological information in
the blockchain through the hash of this public key. However, such
an effort will be herculean for the Central Authority, specially if
it wishes to carry out a mass surveillance of it's citizens. Moreover,
once the pandemic gets over, people will either uninstall the WorkSafeApp
from their smart phone or will keep their bluetooth interface switched
off, which will put an end to any such effort by the authority. 
\end{itemize}

\subsection{Alert Mechanism}

\label{subsec:AlertMechanism} The proposed framework enables the
authority to demarcate an area as a red/orange/green zone. The data
regarding the same (GPS coordinates, radius, date and time of declaration,
etc.), are added as a transaction in the blockchain. Since the blockchain
is publicly available, people will get timely information about the
same through the distributed ledger. Also, an individual who happens
to be in the vicinity of containment zone or is about the enter said
zone, may be alerted through the App.

\subsection{Incentive for Stakeholders}

\label{subsec:Incentive} Success of a framework like this, requires
participation from almost every individual of a Country. However,
it is difficult to realize such a participation if there is no clear
thought on how various sections of the people are going to benefit
from their participation. Apart from the broader social goal of containing
the virus, there has to be perceivable incentive for the individual.
In order to have an idea of how different sections of the society
are going to benefit from this framework, the stakeholders of this
framework are categorized into the following groups, based on their
similarity of interest. 
\begin{itemize}
\item Category-I:Individuals. 
\item category-II: Business houses, Office, shops, Institutes, etc. 
\item Category-III: Testing Centre, Hospitals, Laboratories, police personals. 
\item Category-IV: Authority. 
\end{itemize}
\begin{sidewaystable}
\begin{centering}
\caption{Comparison of CovidChain with other applications and schemes.\label{tab:ComparisonWithOtherApps}}
\par\end{centering}
\begin{tabular*}{0.9\columnwidth}{@{\extracolsep{\fill}}|l|l|c|c|c|c|c|c|c|c|c|c|c|}
\hline 
\textbf{Application} & \textbf{Developer} & \begin{turn}{90}
\textbf{\emph{Anonymity from Other Users}}
\end{turn} & \begin{turn}{90}
\textbf{\emph{Anonymity format Authority}}
\end{turn} & \begin{turn}{90}
\textbf{\emph{Records Epidemiological Data}}
\end{turn} & \begin{turn}{90}
\textbf{\emph{Uses Bluetooth Technology}}
\end{turn} & \begin{turn}{90}
\textbf{\emph{Uses GPS Location Information}}
\end{turn} & \begin{turn}{90}
\textbf{\emph{Records Contact Tracing Data}}
\end{turn} & \begin{turn}{90}
\textbf{\emph{Can be Used as a Digital Pass}}
\end{turn} & \begin{turn}{90}
\textbf{\emph{Has an Alert Mechanism}}
\end{turn} & \begin{turn}{90}
\textbf{\emph{Data Recorded in Distributed Ledger}}
\end{turn} & \begin{turn}{90}
\textbf{\emph{Allows Covid Status Verification}}
\end{turn} & \begin{turn}{90}
\textbf{\emph{Prediction of Infection Risk}}
\end{turn}\tabularnewline
\hline 
TraceTogether \citep{singaporeapp} & Govt. of Singapore & Y & N & N & Y & N & Y & I & I & N & I & N\tabularnewline
\hline 
Aaroyga Setu \citep{aarogyasetu} & Govt. of India & Y & N & Y & Y & Y & Y & Y & I & N & Y & N\tabularnewline
\hline 
Hamagen \citep{israelapp} & Govt. of Israel & Y & N & N & N & Y & Y & N & Y & N & N & N\tabularnewline
\hline 
CovidSafe \citep{covidsafeaustralia} & Govt. of Australia & Y & N & Y & Y & N & Y & I & I & N & Y & N\tabularnewline
\hline 
South Korea & Govt. of South Korea & P & N & Y & Y & Y & Y & I & I & N & I & N\tabularnewline
\hline 
Covid-Watch \citep{covidwatch} & Arx et al. & Y & Y & N & Y & N & Y & N & N & N & N & N\tabularnewline
\hline 
Virusblockchain \citep{virusblockchain} & Hazra et al. & I & I & Y & N & Y & I & P & I & Y & Y & N\tabularnewline
\hline 
WeTrace \citep{DeCarli2020} & Carli et al. & Y & Y & N & Y & Y & Y & I & Y & Y & I & N\tabularnewline
\hline 
Covid-19 BlockChain Framework \citep{Torky2020} & Torky et al. & Y & Y & N & I & I & Y & N & P & Y & I & Y\tabularnewline
\hline 
CovidChain & Choudhury et al. & Y & Y & Y & Y & Y & Y & Y & Y & Y & Y & N\tabularnewline
\hline 
\multicolumn{12}{|l|}{\textbf{Y}: Yes; \textbf{N}: No; \textbf{I}: Insufficient Information;
\textbf{P}: Partially Yes} & \tabularnewline
\hline 
\end{tabular*}
\end{sidewaystable}

Category-I contains two kind of individuals: a) an individual whose covid status is `+ive' or is `In Quarantine (IQ)' b) an individual whose covid status is `Out of Quarantine (OQ)', or an individual who never had any symptoms and never meet an infected person. While people of the first kind would like to see the change of their status (say from `+ive' to `-ive', `-ive' to `IQ' and `IQ' to`OQ', etc.) in the public ledger, people of the second kind would like to carry on       
with their day to day work with minimal restrictions while remaining
protected from infection. If the authority, shops, markets, offices,
businesses, etc., allows access of their facilities only to those
individuals that are found safe, then installation of the WorkSafeApp
will enable people to use their smart phone as a digital-pass to access
these facilities.

Entities in category-II would like to carry on with their day to day
business while taking care of their own and other's safety. By verifying
the covid satus of every individual at entry points, these individuals
will be able to offer a safe environment to their employees and clients.

Entities in Category-III are the ones that generate transactions for
the blockchain by performing clinical test on the individuals. Based
on the previous status of an individual (say `+ive'), in his next
visit, new transaction with status (say `IQ' or `OQ') may be created.
This category will have their obligations to the government and to
the society in general to carry out their responsibilities.

Category-IV will have social obligations to protect the society by
containing the spread of the virus. They are the ones that will be
responsible for setting up such a system and keeping it running. They
will be able to enforce strict quarantine for people who are recorded
as `+ive' or `IQ' in the blockchain. People without a WorkSafeApp
installed smart phone, may not be allowed to work or move freely.
To enforce such restrictions, several countries have adopted the technique
of stamping the hand of quarantined people. However, stamping has its own
limitations like the following. 
\begin{itemize}
\item Stamp impression may fade away before the end of the quarantine period. 
\item Stamp impression may persist even after the quarantine period is over. 
\item People with stamp impression may have to face social stigma. 
\end{itemize}
Another incentive for the Authority to set up such a system is generation
of epidemiological data that may help in finding novel facts and statistics
for containing the virus.

\section{Discussion}

\label{sec:discussion}

In this section, we discuss the different features of the proposed
framework with reference to the applications and proposals discussed
in Section \ref{sec:SimilarWork}. A summary of the comparison of
our proposal `CovidChain' with the other applications and proposals
developed in recent times is presented in Table \ref{tab:ComparisonWithOtherApps}.

In all the related works we have analysed in this area, there is anonymity
from other individuals, but very few have anonymity from the authority
or government. However, in order to gain confidence of the population
so that adoption rate of application increases, ensuring anonymity
of individual from the authority is of paramount important. CovidChain
framework takes care of both the aspects. In CovidChain, an individual
gets to know only the covid status of another individual and though
the authority has access to epidemiological data, it cannot trace
it back to any particular individual.

In many of the works, epidemiological data is not collected. However,
an application like this is possible only if the authority mobilizes
and invests a significant amount of public resources. In return, it
is desirable that new knowledge is derived about the spread pattern
of the virus so that novel strategies for containing the virus may
be planned. Therefore, it is important that epidemiological data is
collected taking care of the anonymity of individuals. In CovidChain,
there is provision for recording of epidemiological data that can
only be accessed by the authority; the general public cannot see the
content as it is encrypted.

Apart form Internet connectivity, two technologies that are commonly
used in these applications are bluetooth and GPS information. While
both these technologies are crucial in exchanging information about
the spread of the virus, it is important that the information that
are transmitted does not lead to compromise in security and anonymity
of an individual. Moreover, one should not be able to transmit misleading
information by impersonating as someone else. In CovidChain, there
is proper mechanism to authenticate an individual. Here, GPS data
related with an individual is never recorded in the blockchain. Only
GPS data related with a containment zone is stored, using which an
individual is alerted by measuring his her distance from a containment
zone.

As evident from the table, the focus of most of the work in this area
is on contact tracing of people who might have come in close contact
of an infected individual. Such works are of great importance as it
may contribute significantly in identifying at-risk individuals who
needs to be urgently quarantined and tested for the virus. However,
in addition to contact tracing, it is equally important to ensure
that quarantined individuals do not put other individuals at risk
by coming out in public. Since, the CovidChain framework allows individuals
to verify each other's covid status, like a digital-pass it may help
in restricting quarantined individuals from accessing public facilities.

In most of the works, accumulated data about the spread of the virus
is stored in a centralized location under the control of a single
authority. For such an arrangement, it is difficult to gain public
support and confidence due to peoples' reservation regarding the privacy
and safety of their data. It becomes difficult to assure people that
their data will not be used for any other purpose after this crisis
is over. Therefore, it becomes imperative that the structure and relevant
portion of the data is publicly made available, so that it is open
for scrutiny and audit. Such an approach not only helps in garnering
public support but also helps various sections of the society to remain
well informed with reliable data that they can use in their individual
businesses. In CovidChain, blockchain technology is used for record
keeping. As a result, same copy of the ledger is maintained at several
locations with immutable data that can be accessed by all sections
of the society.

\section{Conclusion}

\label{sec:Conclusion} To contain the rapid spread of COVID-19, several
smart phone based applications have been developed and proposed in
recent times. Most of the solutions are focused on contact tracing
of people who might have come in close proximity of an infected individual.
However, in the current scenario when continued lock downs have brought
the economy of several countries to the edge, there is a need to explore
technical ways to facilitate both social distancing and economic activities
to go hand in hand. Off late, quite a few countries are contemplating
on allowing a section of their uninfected citizens to use smart phone
applications as digital pass to join their respective work. However,
success of such an approach depend highly on people participation,
which is possible only when data are maintained transparently taking
security and privacy concerns into consideration. In this paper, we
propose a framework that can be used by people to protect themselves
from infections while they are involved in their day to day business
activities. The framework uses blockchain technology for secure and
anonymous record keeping. The distributed nature of blockchain makes
relevant data accessible to all stakeholders for their use and public
scrutiny.

\bibliographystyle{elsarticle-num}
\bibliography{covid}

\end{document}